\documentclass{physeauth}
\usepackage{graphicx}
\usepackage{amsmath}
\usepackage{amssymb}
\usepackage{braket}
\usepackage{empheq}  % fuer die darstellung von Spaltenvektoren u Matrizen
\usepackage{mathrsfs}
\usepackage[usenames]{color}

\newcommand{\oabovea}[2]{\genfrac{}{}{0pt}{3}{#1}{#2}} % zwei terme übereinander schreiben
\newcommand{\dint}[0]{\mbox{d}}                        % integral 'd'
\newcommand{\operator}[1]{\hat{#1}}           % darstellung beliebiger Operatoren

\newcommand{\lagrmulty}[0]{\mathscr{E}}

\newcommand{\conjug}[0]{^{*}}

\begin{document} 

\begin{frontmatter}

\title{Interacting Bosons beyond the Gross-Pitaevskii Mean-Field} 

\author{M. Heimsoth and M. Bonitz}
%\author{M. Bonitz}

\address{Christian-Albrechts Universit\"at zu Kiel, Institut f\"ur Theoretische und Astrophysik, 24098 Kiel, Germany}

% \author[itap]{Martin Heimsoth\thanksref{thank1}},
% \thanks[thank1]{Corresponding author. \\ E-mail: heimsoth@theo-physik.uni-kiel.de}

\begin{abstract}
Systems consisting of cold interacting bosons show interesting collective phenomena such as Bose-Einstein condensation or superfluidity %\cite{filinov_controllingspatial_2008} 
and are currently studied in condensed matter and atomic physics. Of particular interest are nonideal bosons which exhibit strong 
correlation and spatial localization effects.
%There exist several approximation methods to such systems \cite{cederbaum_best_2003,romanovsky_novel_2006,pitaevskii_bose-einstein_2003}. 
Here we analyse the ground-state of a two-dimensional Bose system with a Hartree-Fock type approximation that was first introduced by Romanovsky et al. [Phys. Rev. Lett. {\bf 93}, 230405 (2004)].
 %\cite{romanovsky_novel_2006}. 
We apply this method to a one dimensional system of charged bosons and analyze the behaviour at strong coupling.
\end{abstract}
\end{frontmatter}

%%%%%%%%%%%%

\section{Introduction}
With the first experimental realisation of a Bose-Einstein condensate in 1995 \cite{anderson_observation_1995}, the theory of ultracold bosonic systems became a field of exceptional interest. In the first experiments, the atoms were weakly interacting, and for such systems the Gross-Pitaevskii mean field approximation (GP) is an adequate approach for the analysis of the condensate \cite{dalfovo_theory_1999}. However, recent experiments succeeded in strongly increasing the interaction strengths of the investigated atomic systems \cite{cornish_stable_2000}. With this tuning of the coupling, it was possible, to produce a phase transition from the superfluid to the Mott insulator phase \cite{greiner_quantum_2002}. Also, for charged bosons in traps crystallization and inhomogeneous distribution of the superfluid density has been predicted \cite{filinov_controllingspatial_2008}. For a theoretical description of these effects an approximation beyond GP is required \cite{cederbaum_best_2003,romanovsky_novel_2006,pitaevskii_bose-einstein_2003}.
In this paper we analyse an approximation method for bosonic systems which is based on a Hartree-Fock type factorization and is expected to be valid for a wide range of interaction strength.

In the many-body theory, Hartree-Fock is a standard method to analyse systems of few to many fermions. The related equations can be derived in several different ways. 
One is obtained by expanding the reduced two-particle density matrix in terms of one-particle density matrices in order to obtain an effective single-particle hamiltonian. The resulting solutions to this approach are many-particle states consisting of one single anti-symmetrised product-state. Another common way proceeds 
in the opposite direction by the assumption, that the many-particle state is given by a single anti-symmetrised product state (Slater determinant). By applying the Ritzian principle to this ansatz, one obtains the Hartree-Fock equations.

In contrast to Fermi systems, for bosonic particles the two approaches lead to different systems of equations. Even the ansatz assuming a single Slater permanent can be implemented in various ways differing in the choice of supplementary conditions \cite{cederbaum_best_2003,romanovsky_novel_2006}.

The best known approximation for an ultracold nonideal dilute Bose gas is the Gross-Pitaevskii approximation in which it is assumed, that all particles remain in the same one-particle orbital. Thus the GP equation describes only the condensate fraction of the system without any interaction with the remainig gas.
By definition the GP approximation cannot describe phenomena such as condensate depletion and fragmentation or Mott-insulator phase transitions of cold atomic gases on a lattice. The results become even worse if the interactions increase or if systems with small particle numbers are considered. As we will see later, this approximation cannot describe a number of fundamental properties, such as the localisation of the particles at high coupling strengths.

Another common numerical method to deal with interacting systems is the Configuration Interaction (CI) in wich no approximation, except the limitation of the chosen basis set, is made, e.g. \cite{szabo_modern_1996}. With nowadays computers and highly optimised programs CI can be applied to systems with up to 10 particles.

In this paper we want to analyse an approximation with substantially less numerical effort than CI, that provides high quality numerical results for strongly correlated bosons in traps. This approximation, the Unrestricted bosonic Hartree-Fock method (UBHF), was first introduced by Romanovsky et al. \cite{romanovsky_crystalline_2004,romanovsky_novel_2006,romanovsky_bosonic_2006} and is less restricted than the related ansatz proposed by Cederbaum \cite{cederbaum_best_2003}. While Romanovsky et al. used an explicit analytical ansatz for the single-particle orbitals (discplaced Gaussians) we develop a completely general scheme without any such restriction. This has the advantage that our method is applicable to any interacting Bose system.

Furthermore, our goal is to compare the UBHF results with CI results to obtain a quantitative conclusion about the accuracy of this method. To this end we perform UBHF and CI calculations for $N=2\dots 6$ particles in a harmonic trap using the same basis sets for both cases. Our comparisons concentrates on systems with a coupling parameter in the range from zero to five and shows the excellent quality of this approximation.

\section{Unrestricted bosonic Hartree-Fock ansatz (UBHF)}
The UBHF method is derived from the following ansatz for the many-particle state
\begin{equation}\label{eqn: symmetrised productstate}
 \ket{\Phi}=\ket{12\ldots N}:=\frac{1}{\sqrt{N!}}\sum_{\pi \in S_N} \bigotimes_{s=1}^N\ket{\pi(s)},
\end{equation}
which means that every particle remains in a certain orbital and the underlying many-particle state is a symmetrised product-state. Depending on the imposed additional restrictions, this ansatz contains, as limiting cases, well-known approximations. In particular, the GP approximation is obtained by the additional restriction, that all orbitals are identical \cite{pitaevskii_bose-einstein_2003}, thus the many-particle state is assumed to be totally Bose condensed. A more general approximation would be obtained by the assumption, that two orbitals are either equal or orthogonal to each other. This kind of ansatz has been introduced and analysed extensivly by Cederbaum et al. \cite{cederbaum_best_2003,alon_fragmentation_2005,Cederbaum_time-dependent_2007,streltsov_general_2006}. 

In contrast, the UBHF approximation is the most general case of this Ansatz where {\em no further restriction} to the underlying one-particle orbitals $\ket{1},\ldots,\ket{N}$ is imposed. We only require the many-particle state to be normalised, so the total energy is given by
\begin{equation}
 E = \braket{\Phi|\operator{H}|\Phi}.
\end{equation}
Thus minimising $E$ leads to the functional
\begin{equation}\label{eqn: energyfunctional with condition}
 E(\ket{1}, \ldots, \ket{N}, \lagrmulty) = \braket{\Phi|\operator{H}|\Phi} - \lagrmulty(\braket{\Phi|\Phi}-1),
\end{equation}
where $\lagrmulty$ is a Lagrange multiplyer for the normalisation of $\ket{\Phi}$. The needed equations to determine the one-particle orbitals are obtained by applying the Ritzian principle to this Ansatz via the searched orbitals. This ansatz has been proposed and implemented for the first time by Romanovsky et al. \cite{romanovsky_bosonic_2006,romanovsky_crystalline_2004},
%In his publications, he applied this ansatz to two dimensional Bose-Systems with a particle-number up to $N=8$, but 
however, they used the additional assumption that every orbital is a displaced Gaussian. In the present paper, this restriction will be dropped.

%
%\subsection{Total energy of a symmetrised product state}
%
The expectation value of the energy of a normalised single permanent many particle state with interacting particles is given by \cite{NegeleOrland}
\begin{multline}\label{eqn: Total energy of Slater permanent}
 \braket{\Phi|\operator{H}|\Phi}=\frac{1}{\sqrt{N!}}\sum_{\pi \in S_N}\sum_{l=1}^N\Big(\prod_{s\neq l}\braket{s|\pi(s)}\braket{l|\operator{h}|\pi(l)}\\
+\frac{1}{2}\sum_{k\neq l} \prod_{s\neq k,l}\braket{s|\pi(s)}w_{kl,\pi(k)\,\pi(l)} \Big).
\end{multline}
The norm of such a productstate is proportional to the permanent of the Gramian matrix of the orbitals $\ket{1},\ldots,\ket{N}$:
\begin{equation}
 \|\Phi\|^2 = \frac{1}{\sqrt{N!}}\sum_{\pi \in S_N}\prod_{s=1}^N\braket{s|\pi(s)}.
\end{equation}
It is advantageous to stay in the abstract representation of the orbitals to perform the variational derivative. By applying the rules given in Appendix \ref{chap: Variational derivation of abstract Hilbertspace vectors} one obtains the following equation, by differentiating \eqref{eqn: energyfunctional with condition}
\begin{multline}\label{eqn: determining equations for the orbitals}
\sum_{\pi \in S_N} \Bigg\lbrace P_n\operator{h}\ket{\pi(n)}+\sum_{l\neq n}P_{nl}\braket{l|\operator{h}|\pi(l)}\ket{\pi(n)}\\
 +\sum_{l\neq n}P_{nl} \operator{J}_{l\pi(l)}\ket{\pi(n)} + \frac{1}{2}\sum_{\oabovea{k,l\neq n}{k\neq l}}P_{nlk}w_{kl,\pi(k)\pi(l)}\ket{\pi(n)} \Bigg \rbrace\\ = \lagrmulty \sum_{\pi \in S_N}P_n\ket{\pi(n)}, \quad \text{  for all $n$},
\end{multline}
where we introduced the subscripted entity with a variational number of subscripts
\begin{equation}
P_{i_1\ldots i_n} = P(\pi)_{i_1\ldots i_n} = \prod_{s\neq i_1\ldots i_n}\braket{s|\pi(s)},
\end{equation}
and the doubly subscripted Hartree operator
\begin{equation}\label{eqn: def: Hartree-Operator}
 \operator{J}_{ij}=\int \dint x\,\dint y\, \varphi_i^*(x) w(x,y)\varphi_j(x) \ket{y}\bra{y}.
\end{equation}

By multiplying equation \eqref{eqn: determining equations for the orbitals} from the right with $\bra{n}$, one obtains a closed expression for the Lagrange multiplyer
\begin{equation}
 \lagrmulty=\frac{\braket{\Phi|\operator{H}|\Phi}}{\braket{\Phi|\Phi}},
\end{equation}
i.e $\lagrmulty$ is the total energy of the System. The key equation \eqref{eqn: determining equations for the orbitals} can be solved with a multidimensional minimisation routine without making any assumption on the explicit analytical form of the orbitals.

\section{Numerical results}
We applied the UBHF approximation to few interacting bosons in a one-dimensional harmonic trap described by the hamiltonian
\begin{equation}
 \operator{H}=\sum_i\frac{1}{2}\Big(x_i^2-\frac{\partial^2}{\partial x_i^2}\Big)+\sum_{i< j}\frac{\lambda}{\sqrt{(x_i-x_j)^2+\kappa^2}},
\end{equation}
which has been made dimensionless by using standard oscillator length and energy scales:
the spatial variable $x$ is given in oscillator units $x_0=\sqrt{\hbar/m\omega}$, and $\lambda$ is the ratio of the Coulomb and confinement energy
\begin{equation}
 \lambda=\frac{q^2}{4\pi\epsilon x_0 \hbar \omega}.
\end{equation}
Thus $\lambda=0$ corresponds to an ideal system. The shielding parameter $\kappa$ is needed in $1D$ to make the two-particle integrals convergent. For the limit $\kappa\longrightarrow0$ the system becomes fermionised \cite{girardeau_relationship_1960}. For all calculations in this paper, we choose $\kappa=0.1$. As basis for the single-particle orbitals we chose the eigenfunctions of the ideal system with a total number of basis states of $n_b=15$. We will compare the results with the exact solution obtained with CI \cite{szabo_modern_1996}. The CI calculations are done in exactly the same basis, thus the differences in the results arise exclusively from the UBHF ansatz. 

\subsection{Total energy}
\begin{figure}[]
 \begin{center}
  \includegraphics[angle=0,width=.45\textwidth]{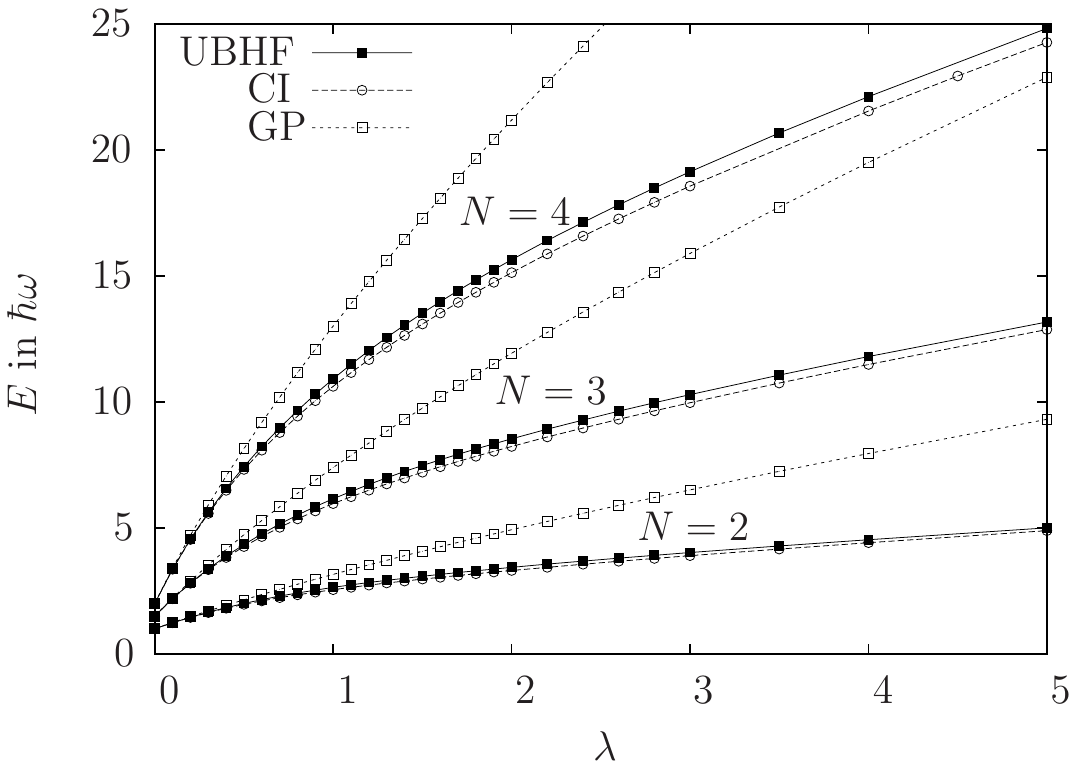}
  \caption{Comparison of the energies obtained with UBHF, GP and CI for different coupling parameters $\lambda$ and for 2,3 and 4 particles.}
  \label{fig: energytest}
 \end{center}
\end{figure}

In Fig.~\ref{fig: energytest} the energies obtained by both methods are compared. Interestingly, the difference between the exact method and the UBHF approximation becomes a constant shift for high interaction values, $\lambda \gtrsim 2$. This shift still depends on the considered particle numbers $N$. As one can see in \ref{fig: energytest}, the shift grows with $N$. Note that for GP the energy diverges rapidly from the exact result already for $\lambda>0.5$. Thus Fig.~\ref{fig: energytest} is a convincing evidence for the good quality of the UBHF approximation.

%In Figs.~\ref{fig: localisation} and \ref{fig: localisation high lambda}, the total particle densities obtained by the different approximation methods are shown. The localisation of the particles due to the interaction is obtained with UBHF, but it is missing within GP. Interestingly, the localisation emerges already in each of the self consistent orbitals ($\varphi_{1/2}$) which are also shown in these pictures.
% Fig.1
% Fig.2
\begin{figure}[]
 \begin{center}
  \includegraphics[angle=0,width=.45\textwidth]{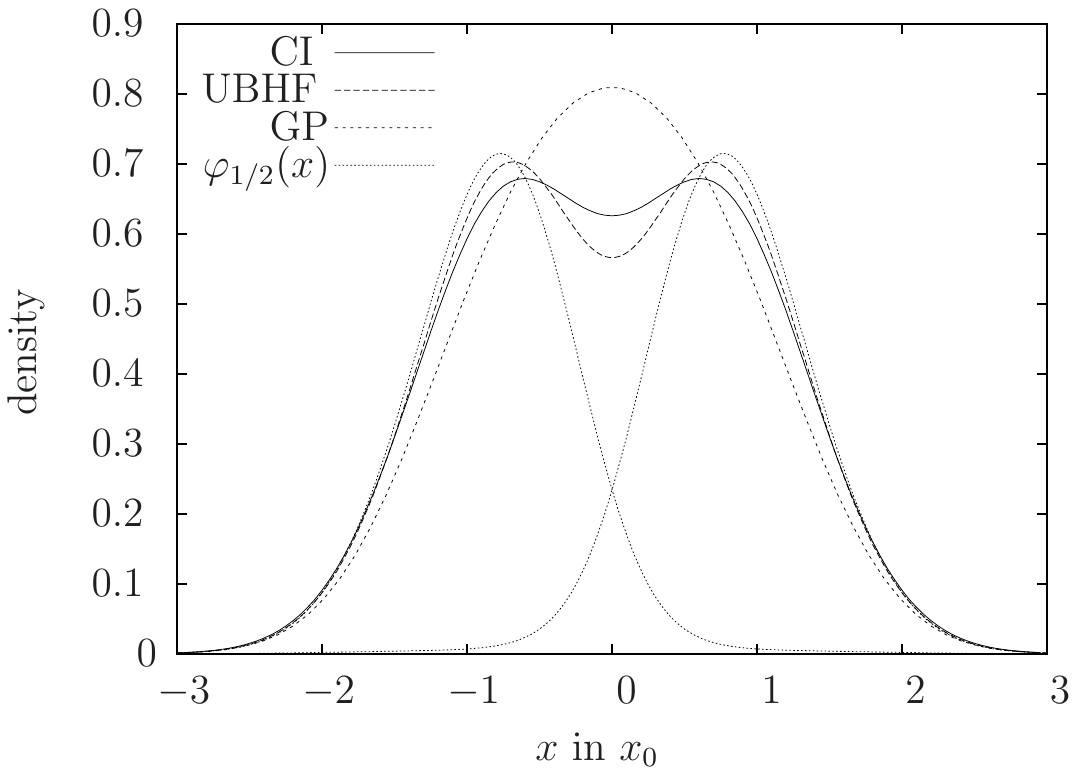}
   \caption{Particle density of a system with $N=2$ and $\lambda=0.9$ for the three different approximation methods. The two orbitals obtained with UBHF are also given.}
  \label{fig: localisation}
 \end{center}
\end{figure}
\begin{figure}[]
 \begin{center}
  \includegraphics[angle=0,width=.45\textwidth]{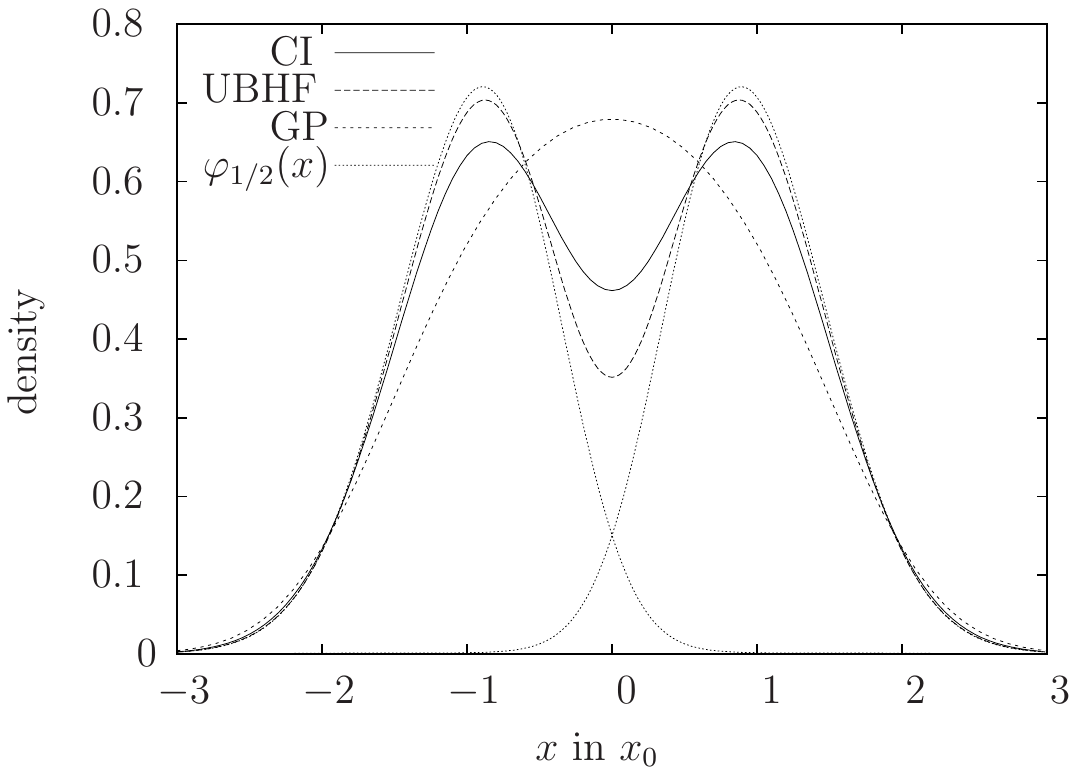}
   \caption{Same as Fig.\ref{fig: localisation}, but for $\lambda$=1.5}
  \label{fig: localisation high lambda}
 \end{center}
\end{figure}

\subsection{Localisation of the orbitals for $\lambda\gg1$}
\begin{figure}[]
 \begin{center}
  \includegraphics[width=.45\textwidth]{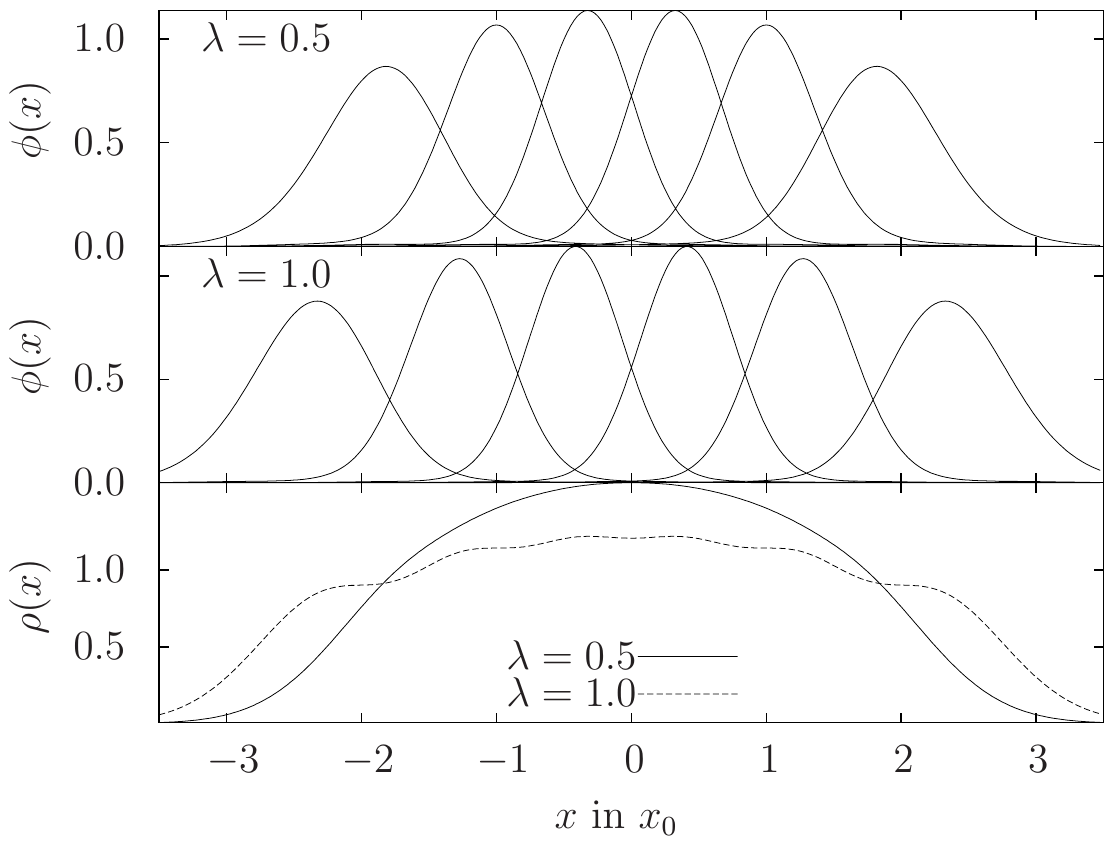}
   \caption{Connection of the obtained orbitals (upper two figures) and the resulting total particle density of a system with $N=6$ particles and $\lambda=0.5$ and $1.0$. }
  \label{fig: delocalisation}
 \end{center}
\end{figure}

The UBHF approximation scheme offers the possibility to analyse the delocalisation of the interacting particles in a special way. As can be seen in Figs.~\ref{fig: localisation}, \ref{fig: localisation high lambda} and \ref{fig: delocalisation}, the overlap of the orbitals vanishes with increasing $\lambda$. It is crucial to notice, that the density from UBHF shows a localisation of the particles which, by definition, is missing in the GP model. Finally, the other two curves in Figs.~\ref{fig: localisation} and \ref{fig: localisation high lambda} which resemble Gaussians are the two orbitals obtained by UBHF. It is interesting to see that the localisation of the particles emerges already in each of the orbitals. This can be understood as a precursor of the classical strongly correlated limit of the system where the particles form a fully localized crystal-like arrangement \cite{afilinov_01prl,bonitz_08pp}. 

This trend is typical for the present system of Coulomb interacting trapped bosons. As another example, in Fig.~\ref{fig: delocalisation} we show the density of six bosons for two values of the coupling parameter. 
Again, with increasing $\lambda$ the overlap of the orbitals decreases and the system, as a whole expands.

\subsection{Delocalisation and non-orthogonality of the orbitals for small $\lambda$}
\begin{figure}[]
 \begin{center}
  \includegraphics[angle=0,width=.45\textwidth]{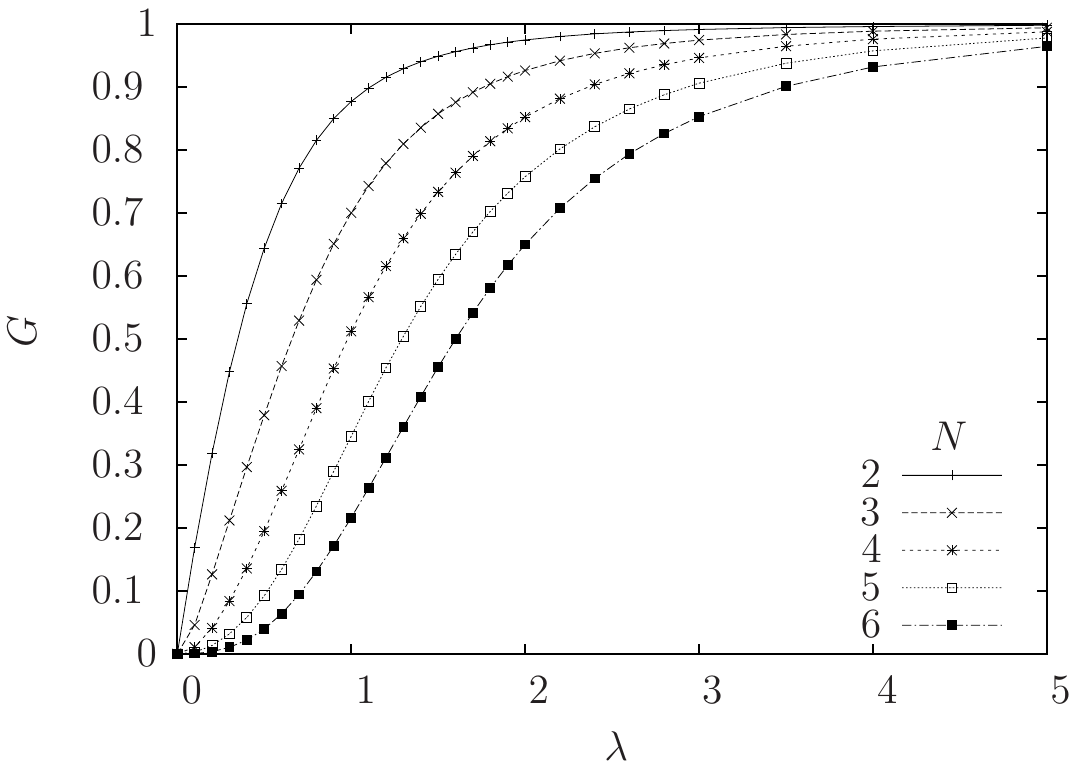}
   \caption{Gramian determinant of the set of orbitals for different particle numbers versus $\lambda$}
  \label{fig: fermionisierung}
 \end{center}
\end{figure}

There are several possibilities to analyse the overlap of orbitals in a quantitative way. One that works for all particle numbers is to consider the Gramian determinant of the given set of orbitals,
\begin{multline}
G(\ket{1},\ldots,\ket{N}) = \sum_{\pi \in S_N}\mbox{sign}(\pi)\prod_{s=1}^N\braket{s|\pi(s)} \\
=\begin{vmatrix}
 \braket{1|1} & \braket{1|2} & \hdotsfor{2} & \braket{1|N} \\
 \braket{2|1} & \braket{2|2} & \hdotsfor{2} & \braket{2|N} \\
 \vdots       & \vdots       & \ddots   &   & \vdots       \\
 \braket{N|1} & \braket{N|2} & \hdotsfor{2} & \braket{N|N} \\
\end{vmatrix}
\end{multline}
This entity is always positive and approaches the value $1$, if the orbitals form an orthonormal set (this is most easily seen for the case of two orbitals). Geometrically, the Gramian determinant is the square of the volume of the parallelepiped spanned by the vectors $\ket{1},\ldots,\ket{N}$ \cite{fischer_lineare_2005}.
In Fig.\ref{fig: fermionisierung} the $\lambda$-dependency of $G$ is shown. Obviously, in the limit of large $\lambda$ the orbitals become pairwise orthogonal. With increasing particle number this limit is reached for larger values of the coupling parameter.

Let us now consider the opposite limit of small coupling. As Fig.~\ref{fig: fermionisierung} shows, for all $N$, the determinant $G$ monotonically decreases when $\lambda$ is reduced. Since $G$ is a determinant it is obvious
that it vanishes, if one or more orbitals are collinear. In the present case, also the opposite is true: vanishing of $G$ is an indication of the orbitals becoming collinear. This is observed for $\lambda \to 0$. Due to the symmetry of the system not only two orbitals but all orbitals are becoming collinear, i.e. they are identical. But this is just the limit of an ideal Bose gas where we expect that all particles Bose condense in the ground state. It is a remarkable property of the present UBHF ansatz which does not impose any restrictions on the orbitals that it yields the correct BEC limit of identical Hartree Fock orbitals for all particles.


% \subsection{Occupation of the natural Orbitals}

\section{Discussion}
%
%\subsection{Complexity of the algorithm}
%
In this paper we have studied a Hartree-Fock approximation for interacting bosons in a trap which has proposed by Romanovsky et al. In contrast to their work we have not used any assumption on the explicit form of the orbitals but obtained them selfconsistently. We have tested the UBHF scheme in detail by comparing the resulting energies and densities with the ones obtained with CI. Due to the identical choice of the basis set for both approximations, the differences in the results arise exclusively due to the approximation made in UBHF. The comparison shows that this is a very accurate model. In contrast, it is confirmed that the GP approximation is not applicable to nonideal charged bosons in traps when the dimensionless interaction strengths $\lambda$ exceeds $0.5$. For its resulting energies are substanially higher than the exact ones and it cannot reproduce important qualitative features of the system such as the localisation of the total particle density. In the limit of vanishing $\lambda$ the UBHF approximation yields single particle orbitals which are collinear -- in other words, we automatically recover the 
GP model.

We mention that the computer time required to solve the equations \eqref{eqn: determining equations for the orbitals}, within our current implementation, has an unfavorable dependence on the particle number $N$ scaling as $N\cdot N!$. This arises from performing the sum over all permutations in  \eqref{eqn: determining equations for the orbitals}. However, it is possible to reduce the calculation of the entities that turn up in this equation to the calculation of permanents. By using the Ryser algorythm, the $N$-dependency of the complexity becomes of order $N\cdot 2^N$ \cite{ryser_permanent_1963}. This is still a rapidly growing calculation time, but it is much better than the first version.
The calculation time also depends on the size of the chosen basis set $n_b$. For both implementations it is of order $n_b^4$. This arises from the calculation of the two-particle-integrals in each step \cite{hurley_Optimized_1988}.

Presently we are working on further optimizing the implementation of this ansatz in order to extend it to systems with higher particle numbers and higher dimensions. Furthermore, the development of an extension to the time dependent regime is in progress.

\section*{Acknowledgements}
We thank K. Balzer for useful discussions.

\appendix
\section{Variational differentiation of abstract Hilbert space vectors}\label{chap: Variational derivation of abstract Hilbertspace vectors}
In order to obtain the key equations for the UBHF approximation, one has to perform a variational derivative with respect to all $N$ orbitals. In this appendix we derive some differentiation rules that are needed to deduce \eqref{eqn: determining equations for the orbitals}. In the following the wavefunction of the $n$-th orbital will be denoted $\varphi_n(x)$. Let us now consider the variational derivative of the matrix element of an arbitrary operator $\operator{O}$:
\begin{multline}
 \frac{\delta}{\delta \varphi^*_n(x)}\braket{i|\operator{O}|j} = \frac{\delta}{\delta \varphi^*_n(x)}\int \varphi^*_i(z)O(x,y)\varphi_j(y) \dint x\dint y \\ = \delta_{in}\int O(x,y)\varphi_j(y) \dint y
 =\delta_{in}\braket{x|\operator{O}|j} \\ = \bra{x}\Big(\frac{\delta}{\delta \bra{n}}\braket{i|\operator{O}|j}\Big),
\end{multline}
where the latter transformation can be regarded as the definition of the derivative with respect to an abstract Hilbert space vector:
\begin{equation}
 \frac{\delta}{\delta \bra{n}}\braket{i|\operator{O}|j}:=\delta_{in}\operator{O}\ket{j}.
\end{equation}
To verify that this definition makes sense we consider the \textit{same} derivative but in an arbitrary representation. Therefore, we differentiate the given matrix element with respect to $c\conjug_{n\gamma}$ -- the $\gamma$-th expansion coefficient of the $n$-th orbital
\begin{multline}
 \frac{\partial}{\partial c^*_{n\gamma}} \braket{\varphi_i|\operator{O}|\varphi_j} = \frac{\partial}{\partial c^*_{n\gamma}} \sum_{\alpha\beta}c^*_{i\alpha}O_{\alpha\beta}c_{j\beta}\\ = \sum_{\alpha\beta}\underbrace{\frac{\partial c^*_{i\alpha}}{\partial c^*_{n\gamma}}}_{=\delta_{in}\delta_{\alpha\gamma}} O_{\alpha\beta}c_{j\beta} = \delta_{in}\sum_{\beta}O_{\gamma\beta}c_{j\beta} \\ = \delta_{in} \braket{\gamma|\operator{O}|j} = \bra{\gamma} \Big(\frac{\delta}{\delta \bra{n}} \braket{i|\operator{O}|j} \Big).
\end{multline}
For the differentiation of the two-particle integrals, we can use the same ideas to derive the following equation
\begin{align}
\frac{\delta}{\delta \bra{n}}(ij|\hat{w}|kl)&=\delta_{in}\operator{J}_{jl}\ket{k}+\delta_{jn}\operator{J}_{ik}\ket{l}\\ \nonumber
 &= \delta_{in}\operator{K}_{jk}\ket{l}+\delta_{jn}\operator{K}_{il}\ket{k}.
\end{align}
Thus we have some freedom to define the remaining operator. The first one that appears in this equation is defined above Eq.~\eqref{eqn: def: Hartree-Operator} whereas the exchange operator $\operator{K}_{jk}$ is given by
\begin{equation}
\operator{K}_{jk}=\int \dint x \,\dint y\, \varphi_j^*(x) w(x,y)\varphi_k(y) \ket{y}\bra{x}.
\end{equation}
With these rules for the functional derivative of matrix elements, together with the product rule,
\begin{multline}
 \frac{\delta}{\delta\bra{n}}\big(F_1[\ldots,\bra{n},\ldots]\cdot F_2[\ldots,\bra{n},\ldots]\big) \\= \Big(\frac{\delta F_1}{\delta\bra{n}}\Big)\cdot F_2 + F_1\cdot \Big(\frac{\delta F_2}{\delta\bra{n}}\Big),
\end{multline}
one derives equations \eqref{eqn: determining equations for the orbitals} by differentiating the functional \eqref{eqn: energyfunctional with condition}.

% \bibliographystyle{unsrt}
% % \bibliographystyle{elsart-num}
% % \bibliographystyle{abbrv}
% % \bibliographystyle{plain}
% 
% \bibliography{bibliography}

\begin{thebibliography}{10}

\bibitem{anderson_observation_1995}
M.~H. Anderson, J.~R. Ensher, M.~R. Matthews, C.~E. Wieman, and E.~A. Cornell.
% \newblock Observation of bose-einstein condensation in a dilute atomic vapor.
\newblock {\em Science}, 269(5221):198, 1995.

\bibitem{dalfovo_theory_1999}
Franco Dalfovo, Stefano Giorgini, Lev~P. Pitaevskii, and Sandro Stringari.
% \newblock Theory of bose-einstein condensation in trapped gases.
\newblock {\em Rev. Mod. Phys.}, 71(3):463, 1999.

\bibitem{cornish_stable_2000}
S.~L. Cornish, N.~R. Claussen, J.~L. Roberts, E.~A. Cornell, and C.~E. Wieman.
% \newblock Stable 85rb bose-einstein condensates with widely tunable interactions.
\newblock {\em Phys. Rev. Lett.}, 85(9):1795, 2000.

\bibitem{greiner_quantum_2002}
Markus Greiner, Olaf Mandel, Tilman Esslinger, Theodor~W. H\"ansch, and Immanuel Bloch.
% \newblock Quantum phase transition from a superfluid to a mott insulator in a gas of ultracold atoms.
\newblock {\em Nature}, 415(6867):39, 2002.


\bibitem{filinov_controllingspatial_2008}
A.~Filinov, J.~B\"oning, M.~Bonitz, and Yu. Lozovik.
\newblock {\em Phys. Rev. B}, 77(21):214527--5, 2008.

\bibitem{cederbaum_best_2003}
L.~S. Cederbaum and A.~I. Streltsov.
\newblock {\em Phys. Lett. A}, 318(6):564, 2003.

\bibitem{romanovsky_novel_2006}
I.~A. Romanovsky.
\newblock PhD thesis, {\em Novel properties of interacting particles in small low-dimensional systems}, Georgia Institute of Technology 2006.

\bibitem{pitaevskii_bose-einstein_2003}
L.~Pitaevskii and S.~Stringari.
\newblock {\em Bose-Einstein Condensation}.
\newblock Oxford University Press, New York, 2003.

\bibitem{szabo_modern_1996}
Attila Szabo and Neil~S. Ostlund.
\newblock {\em Modern Quantum Chemistry: Introduction to Advanced Electronic Structure Theory}.
\newblock Dover Publications, 1996.

\bibitem{romanovsky_crystalline_2004}
Igor Romanovsky, Constantine Yannouleas, and Uzi Landman.
% \newblock Crystalline boson phases in harmonic traps: Beyond the gross-pitaevskii mean field.
\newblock {\em Phys. Rev. Lett.}, 93(23):230405, 2004.

\bibitem{romanovsky_bosonic_2006}
Igor Romanovsky, Constantine Yannouleas, Leslie~O. Baksmaty, and Uzi Landman.
% \newblock Bosonic molecules in rotating traps.
\newblock {\em Phys. Rev. Letters}, 97(9):090401, 2006.

\bibitem{alon_fragmentation_2005}
O.E. Alon, A.I. Streltsov, and L.S. Cederbaum.
% \newblock Fragmentation of bose-einstein condensates in multi-well three-dimensional traps.
\newblock {\em Phys. Lett. A}, 347(1-3):88, 2005.

\bibitem{Cederbaum_time-dependent_2007}
Ofir~E. Alon, Alexej~I. Streltsov, and Lorenz~S. Cederbaum.
% \newblock Time-dependent multi-orbital mean-field for fragmented bose-einstein condensates.
\newblock {\em Phys. Lett. A}, 362(5-6):453, 2007.

\bibitem{streltsov_general_2006}
Alexej~I. Streltsov, Ofir~E. Alon, and Lorenz~S. Cederbaum.
% \newblock General variational many-body theory with complete self-consistency for trapped bosonic systems.
\newblock {\em Phys. Rev. A},
  73(6):063626, 2006.

\bibitem{NegeleOrland}
John~W. Negele and Henri Orland.
\newblock {\em Quantum Many-particle Systems}.
\newblock Westview Press, USA, 1998.

\bibitem{girardeau_relationship_1960}
M.~Girardeau.
% \newblock Relationship between systems of impenetrable bosons and fermions in one dimension.
\newblock {\em J. Math. Phys.}, 1(6):516, 1960.

\bibitem{afilinov_01prl} 
A.~Filinov, M.~Bonitz and Yu.~Lozovik, {\em Phys. Rev. Lett.}, 86:3851, 2001

\bibitem{bonitz_08pp} M. Bonitz et al., {\em Phys. Plasmas}, 15:055704, 2008

\bibitem{ryser_permanent_1963}
H.J. Ryser.
% \newblock Combinatorial mathematics.
\newblock {\em Carus Math. Monograph}, 14, 1963.

\bibitem{hurley_Optimized_1988}
D.L.Huestis J.H.~Hurley and W.A.Goddard.
% \newblock Optimized two-electron-integral transformation procedures for vector-concurrent computer architectures.
\newblock {\em J. Phys. Chem.}, 92:4880, 1988.

\bibitem{fischer_lineare_2005}
G.~Fischer
\newblock {\em Lineare Algebra}.
\newblock Vieweg, Brauschweig/Wiesbaden, 2005

\end{thebibliography}

\end{document}